\documentclass[twocolumn]{aastex63}

\usepackage{enumerate}
\usepackage{xcolor}
\definecolor{forestgreen(web)}{rgb}{0.13, 0.55, 0.13}
\definecolor{ao}{rgb}{0.0, 0.0, 1.0}
\definecolor{lightbrown}{rgb}{0.71, 0.4, 0.11}

\hypersetup{linkcolor=red,citecolor=lightbrown,filecolor=cyan,urlcolor=ao}

\definecolor{forestgreen}{rgb}{0.0, 0.27, 0.13}

\def\be{\begin{equation}}
\def\ee{\end{equation}}
\def\ba{\begin{eqnarray}}
\def\ea{\end{eqnarray}}

\shorttitle{Cosmic rays in Milky Way CGM}
\shortauthors{Jana, Roy, Nath}

\begin{document}

\title{Gamma-ray and radio background constraints on cosmic rays in Milky Way circumgalactic medium}
\correspondingauthor{Ranita Jana}
\email{ranita@rri.res.in}
\author[0000-0003-0436-6555]{Ranita Jana}
\author[0000-0001-9567-8807]{Manami Roy}
\author[0000-0003-1922-9406]{Biman B. Nath}
\affiliation{Raman Research Institute, 
Sadashiva Nagar, 
Bangalore 560080, India}

\begin{abstract}
We study the interaction of cosmic rays (CRs) with the diffuse circumgalactic gas of Milky Way (MW) galaxy that results in hadronic $\gamma-$ray emission and radio synchrotron emission. We aim to constrain the CR population in our circumgalactic medium (CGM) with the help of observed isotropic $\gamma$-ray background (IGRB), its anisotropy and radio continuum. We modify different models of CGM gas in hydrostatic equilibrium discussed in literature by including a cosmic ray population, parametrized by $\eta \equiv P_{\rm CR}/P_{\rm th}$. For the simplest isothermal model, while the IGRB intensity allows $\eta \lesssim 3$, the anisotropy resulting from the Solar system's off-center position in MW rules out all values of $\eta$. For the precipitation model, in which the cooling of the CGM gas is regulated with an optimum ratio of cooling time to free-fall time,while the observed IGRB intensity allows $\eta \lesssim 230$, the observed anisotropy allows only very large values of $\eta$, of order $\gtrsim 100$. The radio continuum limits $\eta \lesssim 400$ for precipitation model and does not constrain isothermal model, however these constraints are mitigated by synchrotron loss time being comparable to CR diffusion time scales. These bounds are relevant for current numerical simulations that indicate a significant CR population in CGM of galaxies of MW mass.
\end{abstract}


\section{Introduction} \label{sec:intro}
Recent numerical simulations have indicated that galactic outflows in Milky Way-type galaxies
can populate the CGM with cosmic rays (CRs). Galactic outflows are likely to contain CR particles, either accelerated in the disk and then advected outwards, or produced by shock acceleration in the outflow. Once these CRs are lifted to the CGM, they would diffuse throughout the halo. Some of the high energy CRs may diffuse out into the intergalactic medium, but most of the CRs would remain in the CGM. For a diffusion coefficient of $D (E) \approx 2\times 10^{29}$ cm$^2$ s$^{-1}$ $E_{\rm GeV} ^{1/3}$ \citep{Berezinsky1997}, and a virial radius of the MW  $\approx 260$ kpc,  CRs with $E\lesssim 1.8$ GeV would be contained in the CGM as their escape time-scale is greater than the age of the Universe. For a shorter and more relevant time scale, the corresponding limit of CR energy would be higher. 

One of the observational implications of having a CR population at large in the CGM is  hadronic interaction of CRs with CGM gas and subsequent $\gamma$-ray production through pion decay. \cite{Feldmann2013} estimated the $\gamma$-ray luminosity of the CGM by solving the transport equation for CRs and assuming a star formation history of MW. They found that the $\gamma$-ray flux from the CGM would provide $\approx 3\%\hbox{--}10\%$ of the total IGRB flux. They did not, however, consider any violent processes such as galactic outflows produced by star formation processes. Similarly, \citet{Liu2019}  used IGRB flux at $\le 1$ TeV to put important limits on CR luminosity ($\le 10^{41}$ erg s$^{-1}$) of MW . In a related simulation, \citet{Chan2019} constrained the average CR diffusivity with observed $\gamma$-ray ($>$ GeV) emission from galaxies. They have found that for dwarf and $L_{\ast}$ galaxies, a constant isotropic diffusion coefficient of order $\sim 3 \times 10^{29}$ cm$^2$ s$^{-1}$ can explain the observed  relation between $\gamma-$ray luminosity and star formation rate. However, they did not compare with synchrotron  observations.

In this {\it Letter}, we ask a related but different question, as to the degree that CRs can dominate the energy budget of the MW CGM, without violating the $\gamma$-ray and radio background limits. This is important  in the context of recent galactic outflow simulations, which depict a picture of the CGM that it may even be dominated by CRs (\citealt{Butsky2018,Dashyan2020,Hopkins2020}). It is also claimed that feedback efficiency of the outflowing gas increases in presence of CRs, by an increase in mass loading and suppression in star formation rate. \citet{Butsky2018} and \citet{Hopkins2020} found that this effect is dependent on the ratio of CR pressure to thermal pressure (which we denote here by $\eta \equiv P_{\rm CR}/P_{\rm th}$) in the CGM. Hence it is necessary to constrain the value of $\eta$ using observational limits. 

For example, while simulating a MW-sized galaxy with different CR transport prescriptions, \citet{Butsky2018} found that $\eta$ can exceed the value 10 over a large portion of the halo, even extending to $\sim 100$ kpc for certain models (see their Fig. 10). \citet{Dashyan2020} simulated smaller galaxies, with virial mass $10^{10}$ and $10^{11}$ $M_\odot$, and found that $\eta$ can have a value $\sim 100$ within central 3 kpc (their Fig. 1). \citet{Ji2019} have found that at redshift $z\lesssim 1$ outflows in MW-mass galaxies can populate the halos with CR and as a result $\eta \approx 10$, although, in warm regions ($T \gtrsim 10^5$ K), locally $\eta$ may have a value less than or comparable to 1.  

We use the IGRB as observed by {\it Fermi-LAT} to constrain the CR population in our halo. While protons in CR population produce $\gamma$-rays, CR electrons in CGM emit synchrotron radiation in the presence of magnetic field. In this regard, we can use the result of \citet{Ravi2013} to constrain CR population who calculated the maximum synchrotron flux that can arise from MW. They showed that a careful modelling of the Galactic components can explain the anisotropic part of the background emission as observed in {\it ARCADE} balloon observations by \citet{Fixsen2011}. This gives an upper limit to the radio frequency emission that can possibly come from CR electron population in an extended halo of our galaxy. We use different density and temperature profiles that have been used in the literature to model the CGM and put bounds on the CR population in the halo.

\section{Density and temperature profiles}
We assume for analytical tractability that CGM gas is in hydrostatic equilibrium in the dark matter potential of the MW. Such models have been recently studied in order to explain the observations of several ions as absorption lines in the lines of sight through the CGM. In order to explore the $\gamma$-ray production implications, we study three illustrative examples of these models: i) Isothermal model (IT),  ii) Precipitation model (PP), and (iii) Isentropic model (IE). 

\begin{figure*}
\includegraphics[width=\textwidth]{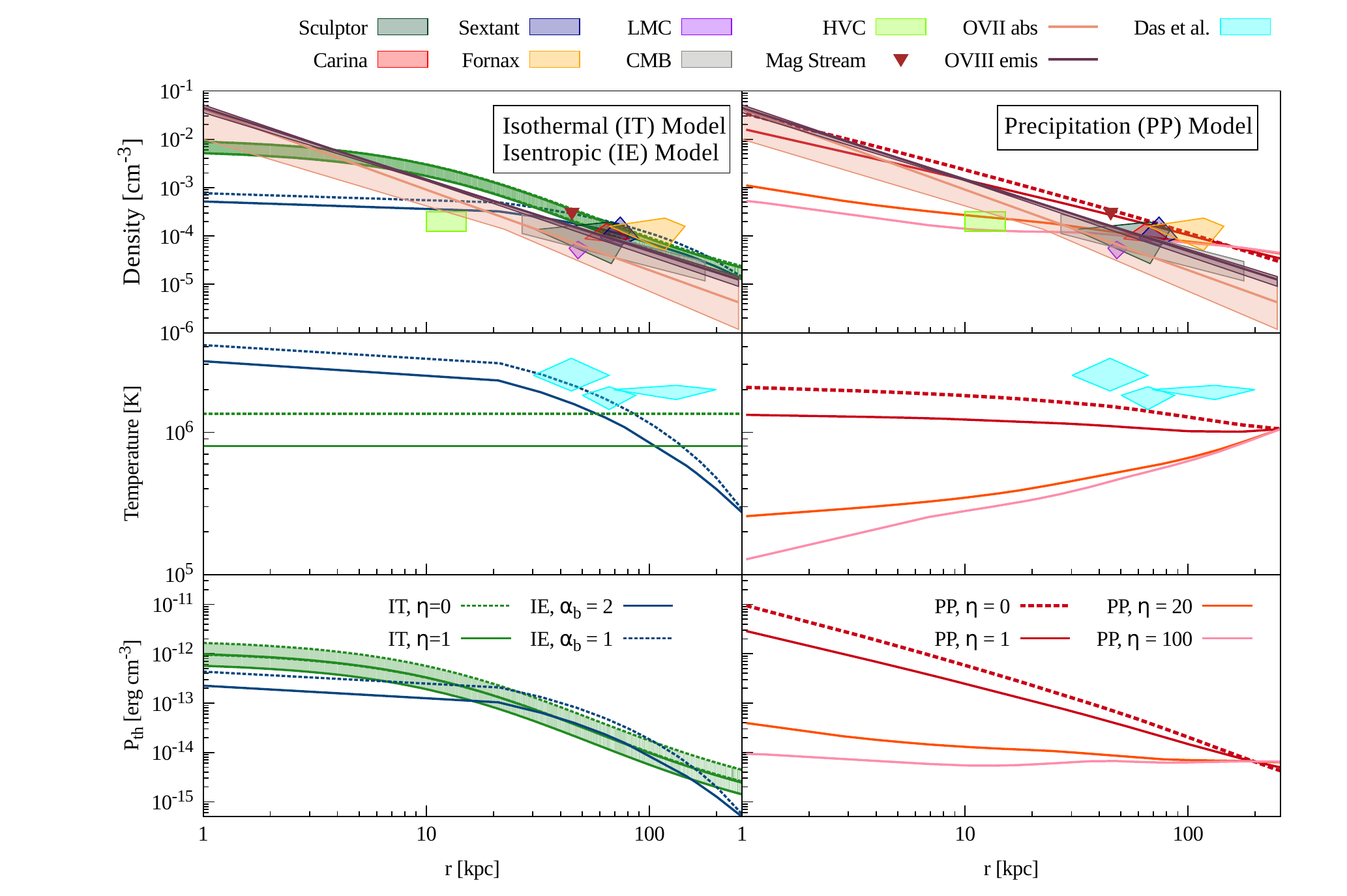}
\caption{Density, temperature and pressure profiles from different models are shown with the distance ($r$) from the Galactic center in the left (Isothermal (IT) and Isentropic (IE)) and right (Precipitation (PP)) panels. IT model is shown for the cases of $\eta=0$ (green, dashed) and $1$ (green, solid)--the density profiles coincide in these two cases, but with two different corresponding temperature and pressure profiles. The no-CR ($\eta=0$) profiles of PP model is shown with dashed red, and those for $\eta=1$ (red solid), $\eta=20$ (orange), $\eta=100$ (pink). The profiles for IE model for $\alpha_{\rm b} =1$ (no-CR, dashed) and $\alpha_{\rm b} =2$ (solid) is shown in blue. Observational constraints are described in detail in text.}
\label{profiles}
\end{figure*}

\begin{figure}
\includegraphics[width=0.45\textwidth]{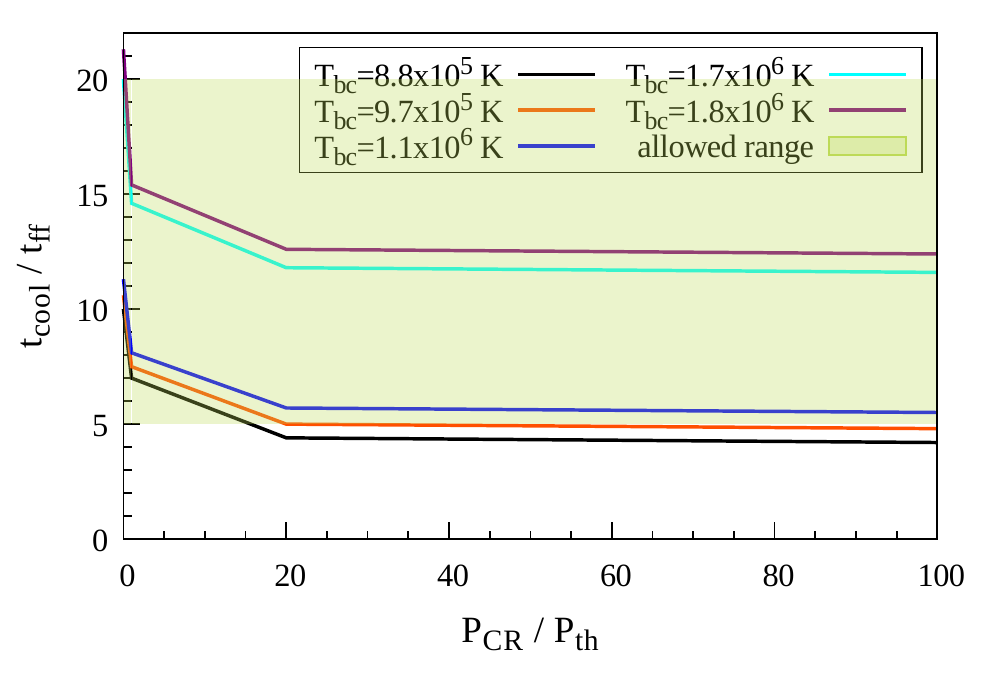}
\caption{Variation of t$_{\rm cool}$/t$_{\rm ff}$ with $P_{\rm CR}/P_{\rm th}$ for different boundary conditions in the precipitation model. The shaded region indicates the allowed range for this ratio of two time scales from cluster studies.}
\label{time_ratio}
\end{figure}

The underlying dark matter potential in all these models is assumed to be that of Navarro-Frenk-White (NFW) profile \citep{Navarro1997}, although in some cases 
we assume a variation of this profile. In the modified version of NFW potential, we assume that the circular velocity $v_c$ is constant ($ = v_{c,max}$) below a radius $2.163 \, r_s$, where $r_s$ is the scaling radius, as suggested by \citet{Voit2019}. We assume a virial mass $M_{\rm vir}=2 \times 10^{12}$ $M_{\odot}$ \citep{Bland2016}, with a concentration parameter $c=10$. When we modify the temperature and density profile by including the non-thermal components, magnetic pressure and CR pressure, we ensure that the total CGM gas mass remains the same. Because of this constraint, the inclusion of a CR population in the CGM  suppresses  gas pressure, by suppressing the gas temperature, as has been also noted in the simulations, e.g, \citep{Hopkins2020}. The Magnetic energy is assumed to be in equipartition with the thermal energy (i.e. $P_{\rm mag} = 0.5\,P_{\rm th}$) in the CGM in absence of any observational constraint. The question of magnetic field strength in the CGM is yet to be observationally settled. On one hand, \citet{Bernet2008} have detected magnetic field in the CGM of galaxies (at $z\sim 1.3$) of comparable strength or larger than that in disks of present-day galaxies. On the other hand, \citet{Prochaska2019} have found in the sightline of a Fast Radio Burst that the magnetic field in the CGM of a massive galaxy is less than the equipartition value. In the absence of any definitive answer, we assume an equipartition magnetic field strength, and calculate the synchrotron flux from CR population in the CGM. In other words, $P_{\text{tot}}=P_{\text{th}}+P_{\text{mag}}+P_{\text{CR}}=P_{\text{th}} (1.5+\eta)$. Below we describe the changes wrought upon by the introduction of CR population in different models.

In the isothermal model, the temperature of the CGM gas is held uniform, and has been extensively used for its simplicity (eg. \citealt{Fang2013}). The  observed temperature of massive halos ($M_{\rm vir}\ge
10^{12}$ $M_\odot$) \citep{Li2015}, and that of MW \citep{Miller2015} is $\ge 2 \times 10^6$ K. In the absence of CR and magnetic field, we assume a uniform CGM temperature of $2\times 10^6$ K. According to the isothermal model of \citet{Miller2015}, the hot gas mass in CGM is within a range of $(2.7\hbox{--}4.7) \times 10^{10}$ $M_\odot$. We therefore initialize our density and temperature profiles such that the CGM contains this amount of mass. In Fig. \ref{profiles}, we show the density, temperature and pressure profiles of IT model with dashed green ($\eta=0$) and solid green ($\eta=1$) lines. The shaded region with the same colour between the dashed (or solid) lines signifies extent of the profiles for a CGM mass within the allowed range for $\eta = 0$ (or 1). The temperature decreases when CR is included, but the density profile practically remains the same, since the CGM mass is held a constant. The temperature falls below the temperature of the photoionized gas ($\sim 10^4$ K) in case of $\eta\geq200$ for this model, hence we only consider $\eta\le 200$ in case of isothermal model. 

In the precipitation model \citep{Voit2019}, the ratio of cooling time to free-fall time ($t_{\rm cool}/t_{\rm ff}$) is assumed to be uniform throughout the halo. 
The underlying idea is that heating and cooling of CGM is regulated in such a manner to keep this ratio at an optimum range. If the ratio becomes too small, cooling would dominate, which would usher in more star formation and stellar feedback would start heating CGM and it would increase the ratio. If the ratio is too large, then reduced feedback would decrease heating, ultimately to pave way for cooling and a reduction of the ratio.
The boundary condition used by \citet{Voit2019} is such that the temperature ($T_{\rm bc}$) at $r_{\rm 200}$ is $kT_{\rm bc}=0.25 \mu m_p v_{c,max}^2$. We use the cooling function $(\Lambda_{\rm N})$ of \textsc{cloudy}, for a metallicity of $Z=0.3Z_\odot$. The total CGM mass in this model for this metallicity is $\approx 6 \times 10^{10}$ $M_\odot$, and we use the same value here. We keep the temperature at the outer boundary ($T_{\rm bc}$, at $r_{200}$) fixed for a particular case when CR is included. Hence the gas temperature in the inner region drops, which increases the cooling rate, and consequently, in order to maintain the same gas mass, the ratio $t_{\rm cool}/t_{\rm ff}$ has to be decreased. According to the simulations for gas in galaxy clusters and massive ellipticals, the optimum range of this ratio is believed to be $5\hbox{--}20$ \citep{Voit2018}. This means that the outer boundary temperature can be varied within a small range, so that this condition is satisfied. We found this range to be $1.1 \times 10^6\hbox{--}1.7\times 10^6$ K, as shown in Fig. \ref{time_ratio}. If the boundary temperature is larger (smaller) than this range, then $t_{\rm cool}/t_{\rm ff}$ becomes larger than $\approx 20$ (smaller than $\approx 5$). We have also included an additional pressure due to turbulence as in the isentropic model, which is described below, and studied its effect on our final results.

The corresponding density, temperature and pressure profiles for PP model are 
shown in Fig. \ref{profiles} with dashed red ($\eta=0$), solid red ($\eta=1$), orange ($\eta=20$) and pink ($\eta=100$) lines. The boundary temperature used for this plot is $1.1\times10^6$ K. The curves show that with an increasing presence of CR, the temperature drops in the inner region, as has also been noted in the simulations of \citet{Ji2019} (their Fig. 5). 

Recently \citet{Faerman2019} have described an `isentropic' model of the CGM, in which entropy is held a constant in the halo. They include three components in their description of pressure: (a) thermal gas (b) non-thermal gas (magnetic field and CR) and (c) turbulence. They characterise turbulence by a fixed $\sigma_{turb}\approx 60$ km s$^{-1}$, and define a parameter $\alpha(r)=(P_{\rm nth}+P_{\rm th})/P_{\rm th}$. They fixed the boundary condition with the help of the value of $\alpha$ at the outer boundary ($r_{\rm 200}$), $\alpha_b$, and varied its value between $1$ (no non-thermal component) and $3$ (equipartition of thermal, magnetic and CR components). In this model, the ratio $\alpha(r)$ drops from its boundary value ($\alpha_{b}$) in the inner region.

In addition to  the density and temperature profiles of these three models, with and without CR, we also show  a few observational constraints on density and temperature in Fig. \ref{profiles} : 
(a)  OVII and OVIII observations \citep{Miller2015}, (b) CMB/X-ray stacking \citep{Singh2018}, (c) limits on density (assuming a temperature of $2.2 \times 10^6$ K from ram pressure stripping of LMC \citep{Salem2015}, Carina, Sextans \citep{Gatto2013}, Fornax, Sculptor \citep{Grcevich2009},(d) pressure equilibrium of high-velocity clouds (assuming the above mentioned temperature) \citep{Putman2012}, and Magellanic stream \citep{Stanimirovic2002}. The observed temperature profile \citep{Das2020} of a $L_\ast$ galaxy NGC 3221 is shown for comparison along with the profiles used here. These constraints show that the density profiles including a CR component are reasonable, although there remains uncertainty regarding the temperature profiles.

\begin{figure*}
\includegraphics[width=\textwidth]{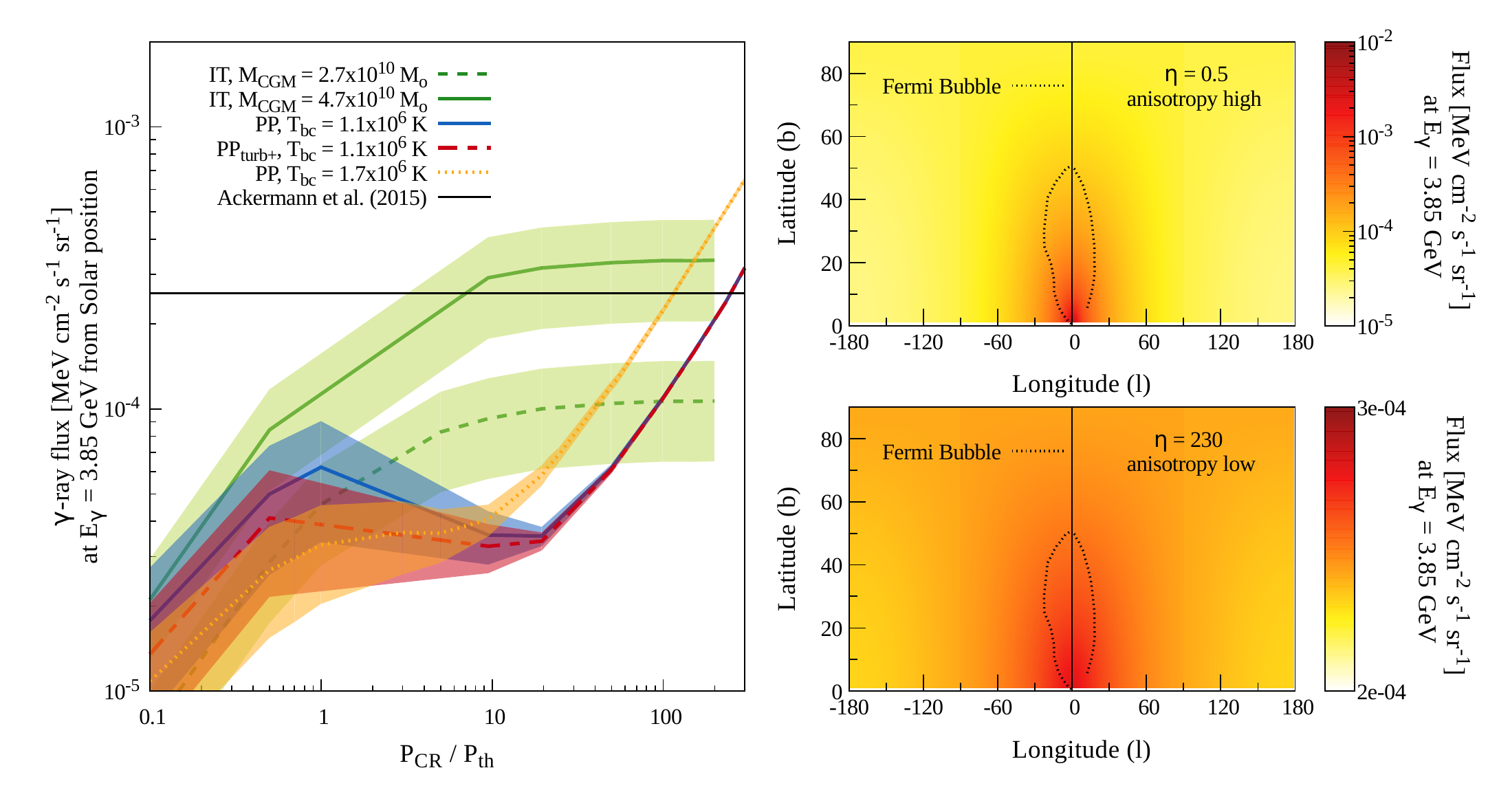}
\caption{{\it Left panel} shows the variation of mean gamma-ray flux from Solar position at 3.85 GeV with $P_{\rm CR}/P_{\rm th}$ for different models and boundary conditions. The black horizontal line shows the  observed flux \citep{Ackermann2015} at $E_{\gamma} = 3.85$ GeV. The curves show the mean flux for $|b| > 30^{\circ}$ and the shaded region around each curve indicates the standard deviation. The case of IT model is shown with a green solid ($M_{\rm CGM} = 4.7 \times 10^{10} M_{\odot}$) and dashed ($M_{\rm CGM} = 2.7 \times 10^{10} M_{\odot}$) line, and PP model with blue solid line ($T_{\rm bc} =$ $1.1 \times 10^6$ K), red dashed line ($T_{\rm bc} =$ $1.1 \times 10^6$ K, with turbulence) and yellow dotted line ($T_{\rm bc} = 1.7 \times 10^6$ K). {\it Right panel} shows the corresponding flux map for PP model ($T_{\rm bc} =$ $1.1 \times 10^6$ K) in Galactic coordinates for $\eta = 0.5$ and $\eta = 230$, made with the angular resolution ($0.6^\circ$) of {\it Fermi-LAT} \citep{Atwood2009}, and in which the dotted line demarcates the region of Fermi Bubble \citep{Su2010}.}
\label{fig:gamma}
\end{figure*}

\begin{figure}
\includegraphics[width=0.5\textwidth]{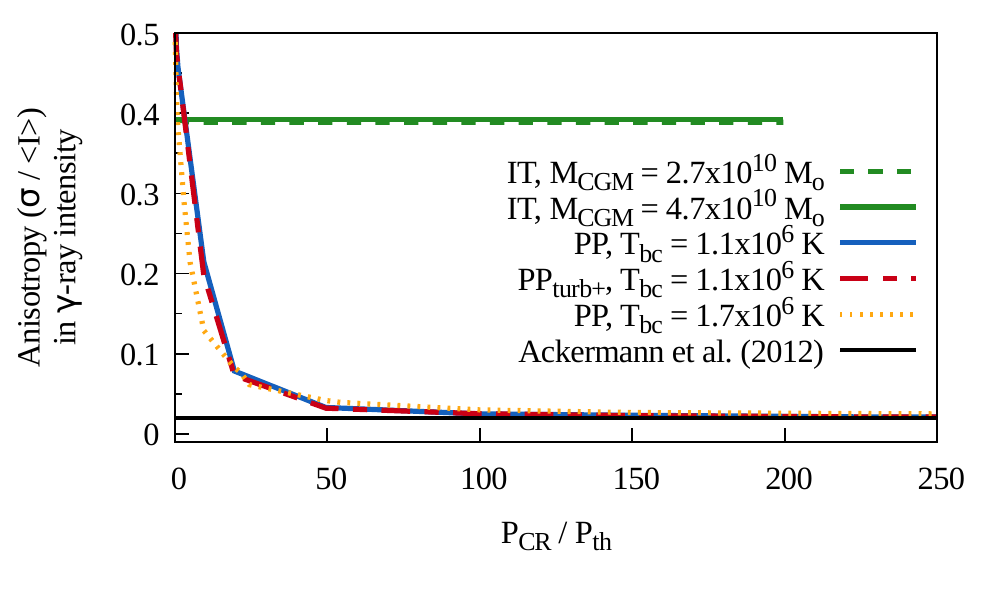}
\caption{Variation of anisotropy (ratio of standard deviation to mean in $\gamma$-ray intensity map) with $\eta$ for different models. The observed value of the ratio (as derived in eqn. \ref{eq:aniso})  \citep{Ackermann2012aniso} is shown with black horizontal line.}
\label{fig:sd}
\end{figure}
\section{Gamma-ray background radiation}\label{sec:gamma}
Once the density and temperature profiles for these models are calculated, we determine the $\gamma$-ray flux resulting due to the hadronic interaction between CR protons and CGM protons. We use the prescription of Dermer's model (\citealt{Dermer1986, Pfrommer2004}) for these calculations. The $\gamma$-ray  flux can be estimated using the source function $\tilde{q}_{\rm \gamma}$ which when multiplied by the number density of target nuclei $(n_{\rm CGM})$, CR energy density ($\epsilon_{\rm CR}$) and photon energy ($E_\gamma$), gives the photon energy per unit time from a particular volume element. The diffuse flux at the Solar position in units of erg cm$^{-2}$ s$^{-1}$ sr$^{-1}$ is then given by

\begin{equation}\label{eq:q_gamma}
J_{\rm \gamma}  =   \int 4 \pi x^2 \, dx\, {1 \over 4 \pi x^2} \, \Bigl [ n_{\rm CGM} (x) \,  \epsilon_{\rm CR} (x) \,  \, E_{\rm \gamma}\, {\tilde{q}_{\rm \gamma}(E_{\rm \gamma}) \over 4 \pi} 
\Bigr ] \,.
\end{equation}

where $x$ is the line-of-sight distance from the position of Solar system. The lower limit ($|b|>30^{\circ}$) of the line of sight integration is chosen in a way such that the contribution from lower latitude, where the Galactic inter-stellar matter dominates over the cicumgalactic medium, is excluded. The omnidirectional source function $\tilde{q}_{\rm \gamma}$ \citep{Gupta2018} is given as, 

\begin{equation}
\tilde{q}_{\rm \gamma}=\left[\frac{ \sigma_{\rm pp} c\,\left(\frac{E_{\pi^{0}}}{\rm GeV}\right)^{-\zeta_{\rm \gamma}} \left[\left(\frac{2E_{\rm \gamma}}{E_{\rm \pi^{0}}}\right)^{\delta_{\rm \gamma}}+\left(\frac{2E_{\rm \gamma}}{E_{\rm \pi^{0}}}\right)^{-\delta_{\rm \gamma}}\right]^{-\zeta_{\gamma}/\delta_{\gamma}}}{ \xi^{\zeta_{\rm \gamma}-2} \left(\frac{3\zeta_{\rm \gamma}}{4}\right) \frac{E_{\rm p}}{2(\zeta_{\rm p}-1)}\left(\frac{E_{\rm p}}{\rm GeV}\right)^{1-\zeta_{\rm p}}{\rm B}(\frac{\zeta_{\rm p}-2}{2},\frac{3-\zeta_{\rm p}}{2})}\right]\ .
\end{equation}
Here $\xi=2$ is the multiplicity factor, $E_{\rm p}$ and $E_{\rm \pi^{0}}$ are the rest mass energy of protons and pions ($\pi^{\rm 0}$), $\zeta_{\rm p}$ and $\zeta_{\rm \gamma}$ are the spectral indices of the incident CR protons and emitted $\gamma$-ray photons respectively, $\delta_{\gamma} = 0.14\zeta_{\rm \gamma}^{-1.6} + 0.44$ is the spectral shape parameter, $ \sigma_{\rm pp}=32(0.96+e^{4.4-2.4\zeta_{\rm \gamma}})$ mbarn (see Equations (8), (19)-(21) in \citealt{Pfrommer2004}) and B stands for beta function. We use $\zeta_p=\zeta_\gamma=2.3$ in our calculations following the spectral fit of \citet{Ackermann2015}.

CR electrons can also produce GeV $\gamma$-ray flux by boosting CMB photons via inverse Compton scattering.  Such electrons  will have TeV range energy. The inverse Compton loss time scale of these high energy electrons is short, $t_{\rm comp} \approx 1.2 \, {\rm Myr} \, (\rm GeV/E_{\gamma})^{1/2}$ where $E_{\gamma}$ is the scattered $\gamma$-ray energy.  In light of this short time scale, we do not consider leptonic process here.

We choose the energy band of $3.2\hbox{--}4.5$ GeV as a representative band for our comparison of model fluxes with observations since the {\it Fermi-LAT} spectral fit of IGRB with index $-2.3$ fits well the data in this band. We compute fluxes at the midpoint of this band $3.85$ GeV for different models and compare with observed flux in the band. 

The $\gamma$-ray flux scales as $\epsilon_{\rm CR} n \propto (\eta \times n^2 T)$, an increase in $\eta$ suppresses the thermal pressure, so the resultant flux depends on the competition between $\eta$ and $n^2T$ terms. For the isothermal model, the more CR there is in CGM, the lower is the gas temperature, but the density profile remains approximately unchanged. This makes the $\gamma$-ray flux increase with the increase in $\eta$. For higher values of $\eta$ (i.e. $\eta \gtrsim 10$ ) the curve flattens because the increase in $\eta$ is compensated by the decrease in temperature (flux $\propto \eta T$ for IT model). 

The case of PP model is interesting, since the density profile is coupled to the temperature and cooling function by $n\propto T(r)^{\frac{3}{2}}/\Lambda_{\rm N}(T(r))$. The initial rise of $\gamma$-ray flux with increasing $\eta$ results from the fact that the temperature is in a range where the cooling function has a plateau and the density profile does not change with $\eta$, but the $\gamma$-ray flux does. This is followed by a decrease in the flux when the temperature is lowered further, and the steep portion of the cooling function suppresses the density, decreasing the $\gamma$-ray flux. For larger $\eta$, the density profile becomes almost flat and any further increase in $\eta$ increases the $\gamma$-ray flux.

The anisotropy in IGRB can also give additional bounds on $\eta$. The fluctuation in IGRB intensity can be decomposed in spherical harmonics as ${\delta I (\theta) \over \langle I \rangle}=\sum_{l,m} \, a_{l,m} Y_{l,m} (\theta)$, where $\delta I(\theta)=I(\theta)-\langle I \rangle$ is the difference in intensity between the mean intensity and the intensity in direction $\theta$. With $C_l=\langle \vert a_{lm} \vert ^2 \rangle$, the correlation function between lines of sight related through $\mathbf{k_1}\cdot\mathbf{k_2}=\cos \theta$ is given by,
\be
C(\theta)=\langle {\delta I({\mathbf k_1}) \over \langle I \rangle} {\delta I ({\mathbf k_2}) \over \langle I \rangle} \rangle=\sum_l \, {2l+1\over 4\pi} C_l \, P_l (\cos \theta)\,.
\ee
Since the Legendre polynomials $P_l(1)=1$, we have from the auto-correlation ($C(\theta=0)$), the ratio of standard deviation to mean intensity, 
\be
{\sigma \over \langle I \rangle}= \Bigl ( \sum_l \, {2l+1\over 4\pi} {C_l \over \langle I \rangle^2} \Bigr ) ^{1/2} \approx 0.02 \,,
\label{eq:aniso}
\ee
where the sum is dominated by $C_l$ at $l=30$ \citep{Ackermann2012aniso}.
   
We show in the right panel of Fig. \ref{fig:gamma} two simulated maps in Galactic coordinates for $\gamma$-ray intensity at 3.85 GeV for $\eta=0.5$ and $\eta=230$ of precipitation model ($T_{\rm bc} =$ $1.1 \times 10^6$ K), made with Fermi-LAT angular resolution of $0.6^\circ$ at 3.85 GeV. The ratio of standard deviation to mean intensity for $|b|>30^\circ$ as a function of $\eta$ is shown in Fig. \ref{fig:sd} for different models. For PP model, the decrease in anisotropy with the increasing $\eta$ results from the flattened out density and temperature profiles. In contrast, anisotropy does not change with $\eta$ for IT model due to unchanged density profile.

The above discussion leads us to two constraints on the CR population in CGM. Firstly, if we consider a 1$\sigma$ spread around the mean intensity, then we get a limit from the observed intensity itself, ruling out those values of $\eta$ for which the intensity (mean+1$\sigma$) exceeds the observed value. This leads to an upper limit of $\eta\lesssim 3$ for IT model, and  $\eta \lesssim 230$ for PP model. Secondly, one can limit $\eta$  considering the  anisotropy, requiring the ratio of standard deviation to mean intensity to be $\le 0.02$. This rules out all varieties of IT models. For PP model, the anisotropy asymptotically reaches the observed limit for large values of $\eta$ ($ \gtrsim 100$). Hence, one can conclude that IGRB intensity and anisotropy allow $100\lesssim \eta \lesssim 230$  for PP model.

The isentropic model has to be dealt separately, since their model already predicts a non-thermal component in its profile. In order to put a limit, we have not included any magnetic pressure and assume $P_{\rm CR}=P_{\rm nth}$ as their model does not allow equipartition of magnetic field in the inner region of halo for the $\alpha_b<4$, and  calculate the corresponding $\gamma$-ray flux at $3.85$ GeV, as a function of the boundary value ($\alpha_b$) of their model. We find that for the isentropic model, the $\gamma$-ray flux never exceeds the {\it Fermi-LAT} data, and at the most has a value $\sim 10\%$ of the flux as this model does not admit a CR dominated CGM in inner region of halo.

\begin{figure}
\includegraphics[width=0.5\textwidth]{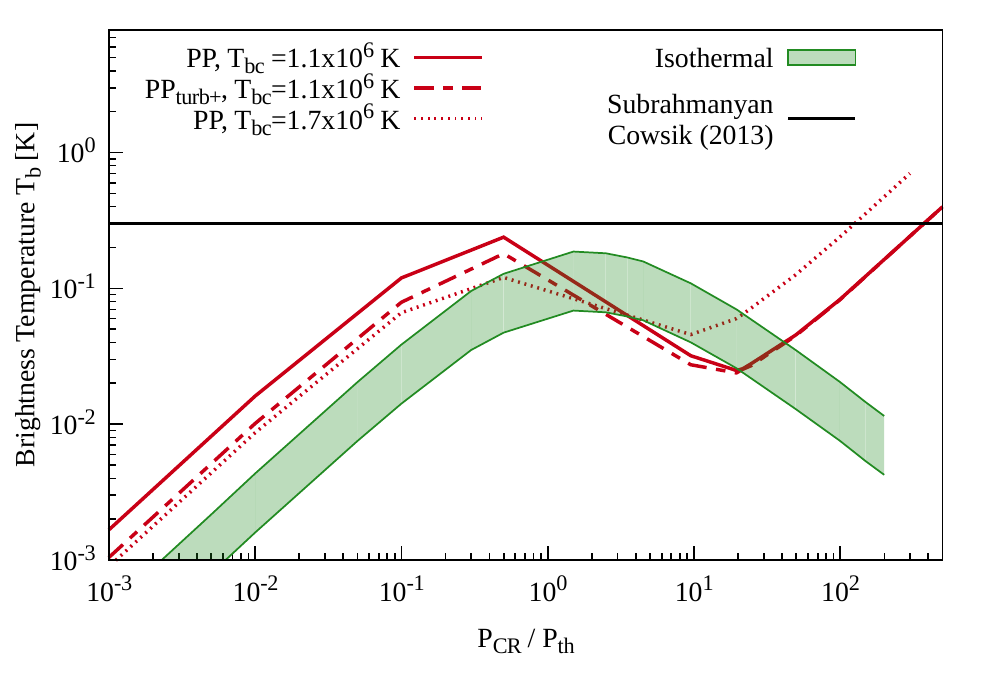}
\caption{The (solid, dashed and dotted red) curves show the brightness temperature at 1.4 GHz for precipitation model for different boundary conditions, and the green band shows the same for isothermal model for the range of CGM mass mentioned in the text. The horizontal black line is the brightness temperature from the halo model of \cite{Ravi2013}.
}
\label{fig:radio}
\end{figure}

\section{Synchrotron radiation}\label{sec:radio}
CR electrons radiate synchrotron emission in the presence of magnetic field. We assume an equipartition magnetic field in the CGM for our calculation, since its value is still a debatable issue. We take the fiducial value of the ratio of CR electrons to protons energy to be $0.01$. Its value is rather uncertain, both theoretically and observationally. From observations in the Solar system, at CR energy $\sim 10$ GeV, where solar modulation effects are low, the ratio is known to be $1\%$.

We assume that the CR electrons have a power-law energy distribution, with the same power-law index $\zeta_p$ as for protons. The observed radio spectrum has an index of $-0.599 \pm 0.036$ (Table 6 of \cite{Fixsen2011}), which would imply $\zeta_p \sim 2.2$ which is not very different from our assumed value. The corresponding radio flux can be calculated using the emissivity (eqn. 6.36 of \citet{Rybicki2004}) and then performing a similar integral as in the case of $\gamma$-ray flux. Finally the brightness temperature is calculated at $1.4$ GHz, in order to compare with observations.

As explained earlier, \citet{Ravi2013} devised a model of the MW synchrotron emitting halo in such a way as to explain the observed radio background towards the Galactic pole, by {\it ARCADE-2}. The purpose of the model of \cite{Ravi2013} was to maximally explain the radio observations with the help of MW halo. This particular model, therefore, gives the maximum possible radio continuum emission that can be attributed to MW halo, and becomes useful for our purpose of putting limits on CR electrons in CGM.
 
We show the comparison of synchrotron flux from different models as a function of $\eta$, with the observed limit, in Fig. \ref{fig:radio}. The trends of radio flux with $\eta$ are different from the case of $\gamma$-ray, because here the magnetic field is pegged to the thermal pressure. We find that in the isothermal model all values of $\eta$ are allowed. Although in the precipitation model only $\eta \lesssim 400$ keeps the brightness temperature within limit.

The magnetic field in different models range between $(0.2\hbox{--}10)\mu$G (from outer to inner regions), for $\eta=1$. For higher values of $\eta$ the range would be lower. The synchrotron loss time of electrons (responsible for radiating at $1.4$ GHz, with energy $\approx 17.4 \, B_{\mu G}^{-1/2}$ GeV) is $\approx 700 \, {\rm Myr} \, B_{\mu G}^{-3/2}$. The diffusion time scale for the CR electrons to cross 50 kpc radius is $\approx 630 \, {\rm Myr} \, E_{\rm GeV}^{-1/3} \approx 243\, \rm Myr\, B_{\mu G}^{-1/6} $. For low values of $\eta$ ($\eta \sim 1$) most of the contribution to the radio flux comes from within $50$ kpc, hence a  spectral break at $1.4$ GHz is not expected for lower values of $\eta$. For higher values of $\eta$ ($\eta \sim 100$) a spectral break at 1.4 GHz will appear at $\sim 2$ Gyr (synchrotron loss time) when CR diffuses beyond $\sim$ 150 kpc from where half of the radio emission occurs. This will decrease the radio flux for large $\eta$, which should be noted with regard to our limits on $\eta$ above.

\section{Discussions}
The variations of radio and $\gamma$-ray fluxes with $\eta$ for different  boundary conditions in Fig. \ref{fig:gamma} and \ref{fig:radio} show that our constraints are rather robust. We also show the result of inclusion of turbulence support in the CGM (red dashed lines), which indicate, again, the robustness of our constraints. However, it is possible that non-linear processes such as CR streaming instability may change the density profile \citep{Rus2017} and change the conclusions.

We note that the $\gamma$-ray and radio flux, hence the limit of $\eta$, depend on CGM gas mass. A $10\%$ increase (decrease) in CGM mass would result $\lesssim 30\%$ increase (decrease) in both the fluxes.

The limit on CR electrons through synchrotron emission depends on the assumption of equipartition strength of magnetic field. If the magnetic field strength were to be  a fraction $\psi$ of the equipartition value, then the synchrotron flux would scale as $\propto \psi^{(\zeta_p+1)/2}$. For $\psi=0.1$, {\it e.g,} the flux would decrease by a factor $0.02$, for $\zeta_p=2.3$ considered here, thereby making the synchrotron limits on $\eta$ practically irrelevant. 

\section{Summary}
We have pointed out that IGRB  and radio continuum background  can act as important checks for models that populate CGM with a significant amount of CR. Using various density and temperature profiles from literature we have shown that resulting $\gamma$-ray background and the associated anisotropy constrain the CR pressure to thermal pressure ratio $100 \lesssim \eta (\equiv P_{\rm CR}/P_{\rm th}) \lesssim 230$ in the precipitation model, the lower limit arising from anisotropy due to the off-center position of the Solar system in MW, and the upper limit, from IGRB intensity measurements. Although the isothermal model allows $\eta \lesssim 3$ considering the intensity (mean + 1$\sigma$), but anisotropy considerations rule out all values of $\eta$ in this model. Limits from radio background ($\eta \lesssim 400$ for precipitation model) are rather weak in comparison.

\section*{Acknowledgements}
We would like to thank Kartick Chandra Sarkar, Sayan Biswas, Shiv Sethi and Prateek Sharma for valuable discussions and an anonymous referee for detailed comments.

\bibliography{reference.bib}{}
\bibliographystyle{aasjournal}

\end{document}